\documentclass{article}
\usepackage{spconf,amsmath,graphicx, svg, longtable, enumitem}
\usepackage{CJKutf8}

\usepackage{multirow,hhline,graphicx,array}
\graphicspath{{fig/}}

\usepackage[font=small,labelfont=bf]{caption}
\usepackage{hyperref}
\usepackage{cleveref}
\title{AUDIO CAPTION: LISTEN AND TELL}
\name{Mengyue Wu, Heinrich Dinkel and Kai Yu\thanks{Heinrich Dinkel is the co-first author. Kai Yu and Mengyue Wu are the corresponding authors. 
This work has been supported by the Major Program of National Social Science Foundation of China (No.18ZDA293). Experiments have been carried out on the PI supercomputer at Shanghai Jiao Tong University.}}
\address{MoE Key Lab of Artificial Intelligence\\
	SpeechLab, Department of Computer Science and Engineering\\
    Shanghai Jiao Tong University, Shanghai, China\\
    \texttt{\small\{mengyuewu, richman, kai.yu\}@sjtu.edu.cn}
}

\begin{document}

\begin{CJK*}{UTF8}{gbsn}
\small
\maketitle
\begin{abstract}
Increasing amount of research has shed light on machine perception of audio events, most of which concerns detection and classification tasks. However, human-like perception of audio scenes involves not only detecting and classifying audio sounds, but also summarizing the relationship between different audio events. Comparable research such as image caption has been conducted, yet the audio field is still quite barren. This paper introduces a manually-annotated dataset for audio caption. The purpose is to automatically generate natural sentences for audio scene description and to bridge the gap between machine perception of audio and image. The whole dataset is labelled in Mandarin and we also include translated English annotations. A baseline encoder-decoder model is provided for both English and Mandarin. Similar BLEU scores are derived for both languages: our model can generate understandable and data-related captions based on the dataset. 
\end{abstract}
\begin{keywords}
Audio Caption, Audio Databases, Natural Language Generation, Recurrent Neural Networks
\end{keywords}
\section{Introduction}
\label{sec:intro}

Current audio databases for audio perception (e.g. AudioSet \cite{45857}, TUT Acoustic Scenes database \cite{Mesaros2016TUTDF}, UrbanSound dataset \cite{Salamon2014ADA}) are mainly segmented and annotated with individual labels, either free of choice or fixed. These datasets provide a systematic paradigm of sound taxonomy and generate relatively great results for both classification \cite{xu2018large} and detection tasks \cite{mcfee2018adaptive}. However, audio scenes in real world usually consist of multiple overlapped sound events. When humans perceive and describe an audio scene, we not only detect and classify audio sounds, but more importantly, we figure out the inner relationship between individual sounds and summarize in natural language. This is a superior ability and consequently more challenging for machine perception. 

Similar phenomenon is present with images, that to describe an image requires more than classification and object recognition. To achieve human-like perception, using natural language to describe images(\cite{Vinyals_2015, Karpathy:2017:DVA:3069214.3069250, xu2015show}) and videos has attracted much attention(\cite{Shen_2017, Pasunuru_2017, Pan_2016}). Yet only little research has been made regarding audio scenes\cite{8170058}, which we think is due to the difference between visual and auditory perception. For visual perception, spatial information is processed and and we could describe the visual object by its shape, colour, size, and its position to other object. However for audio sounds, the traits are to be established. Auditory perception mainly involves temporal information processing and the overlap of multiple sound events is the norm. To describe an audio scene therefore requires a large amount of cognitive computation. A preliminary step is to discriminate the foreground and background sound events, and to process the relationship of different sounds in temporal order. Secondly, we need to acquire our common knowledge to fully understand each sound event. For instance, a 3-year-old child might not entirely understand the meaning of a siren - they could not infer that an ambulance/fire engine is coming. Our common knowledge accumulates as we age. Lastly, for most audio scenes involving speech, we need access to the semantic information behind speech signals to fully understand the whole audio scene. Sometimes further reasoning based on the speech information is also needed. For example, through a conversation concerning treatment option, we could speculate this might be between a doctor and a patient. 

Nevertheless, current machine perception tasks are mostly classification and detection. In order to help machine understand audio events in a more human-like way, we are in need of a dataset that enables automatic audio caption. Its aim is to automatically generate natural language to describe an audio scene. Broad application can be expected: hearing-impaired people can understand the content of an audio scene and detailed audio surveillance will be possible. Just as humans need both audio and visual information for comprehensive understanding, the combination of these two channels for machine perception is inevitable. It is therefore practical and essential to have a tool to help process audio information. Further, English is quite a dominant language in captioning field. Among the limited attempts to caption images in other languages, a translation method is usually used (e.g. \cite{Li2016}) and to our knowledge, there is no audio dataset specifically set up for Mandarin caption. 

Section 2 will introduce the Audio Caption Dataset and detailed data analysis can be found in Section 3. Baseline model description is provided in Section 4. We present human evaluation results of the model generated sentences in Section 5. 

\section{The AudioCaption Dataset}
\label{sec:format}

Previous datasets (e.g. AudioSet \cite{45857}) mostly concentrate on individual sound class with single-word labels like music, speech, vehicle etc.. However these labels are insufficient to probe the relationship between sound classes. For instance, provided with labels `speech' and `music' for one sound clip, the exact content of an audio remains unclear. Further, although AudioSet contains as many as 527 sound labels, it still cannot include all sound classes, especially some scene-specific sounds. 

This database departs itself from previous audio datasets in four aspects 1)the composition of more complicated sound events in one audio clip; 2) a new annotation method to enable audio caption; 3) the segmentation of specific scenes for scene-specific audio processing; 4) the use of Mandarin Chinese as the natural language. We also include translated English annotations for broader use of this dataset. We identify five scenes that might be in most interest of audio caption - hospital, banking ATMs, car, home and conference room. We firstly reveal our 10h labelled dataset on hospital scene and will keep publishing other scenes. 

\paragraph*{Source} As audio-only information is not sufficient for determining a sound event, we included video clips with sound. All video data were extracted from Youku, Iqiyi and Tencent movies which are video websites equivalent to Youtube. They also have exclusive authorization of TV shows, thus some of the video clips were from TV shows and interviews. When sampling the data, we limited background music and maximized real-life similarity. Each video was 10s in length, with no segmentation of sound classes. Thus our sound clips contain richer information than current mainstream datasets. The hospital scene consists 3710 video clips (about 10h duration in total, see \Cref{tab:data_stats}).

\paragraph*{Annotation} We think of four traits to describe the sound evetns in an audio clip: its definition (what sound is it), its owner (what's making the sound), its attribute (how does it sound like), its location (where is the sound happening). Almost every event scene can be understood via these four traits, e.g. ``Two dogs are barking acutely'',``A bunch of firemen are putting out fire and there are people screaming''. 

Each video was labelled by three raters to ensure some variance of the dataset. All human raters received undergraduate education and were instructed to focus on the sound of the video while labelling. They were asked to label in two steps:\\

\vspace{-5mm}
\begin{enumerate}[label*=\arabic*.,noitemsep, topsep=0pt]
     \item Answer the following four questions:
    \begin{enumerate}[label=\arabic*), noitemsep]
        \item list all the sounds you’ve heard; 
        \item who/what is making these sounds;
        \item how are these sounds like; 
        \item where is the sound happening. 
        \end{enumerate}
    \item Use natural language to describe the audio events;
\end{enumerate}

The questions in Step 1 are to help generate a sentence like the description in Step 2, which are the references in our current task. All the labelling language was chosen freely as there is great subjective variability in human perception of the audio scene.

\section{Data Analysis}


Our raters used free Mandarin Chinese to label the dataset, a character-based language in which several characters can signify the same meaning. For instance, ‘声音’ ‘声’ ‘声响’ all mean sound. ‘音乐’ means music, and ‘音乐声’ ‘音乐声音‘ mean the sound of music. Consequently, we firstly need to merge those labels with the same meaning. \Cref{tab:sound_label} is a showcase of the most frequent 15 sound labels in our dataset after merging, collected from the answers to Question 1. Most frequent sound belongings and scenes can be seen in \Cref{tab:sound_scene}. All labelling results were then translated into English by Baidu Translation to increase the accessibility of the dataset. 

\begin{table}[tb]
    \centering
    \small
        \begin{tabular}{c|l|l}
         \hline
         Rank & Tags & Instances \\ [0.5ex]
         \hline
         \hline
         1 & People speaking/talking & 24.70\% \\
         2 & People having conversation & 7.42\% \\
         3 & Footsteps/Walking & 5.65\% \\
         4 & Post production sound & 5.34\% \\
         5 & Medical machine sound & 3.96\% \\
         6 & Noisy background sound & 2.68\% \\
         7 & Music & 1.23\% \\
         8 & Monologue & 0.95\% \\
         9 & Crying & 0.69\% \\
         10 & Door opening/closing & 0.54\% \\
         11 & Medical instrument sound & 0.43\% \\
         12 & Laughing & 0.43\% \\
         13 & Friction & 0.28\% \\
         14 & Coughing & 0.26\% \\
         15 & Collision & 0.24\% \\
         \hline
    \end{tabular}
    \caption{Top 15 Sound Tags }
    \label{tab:sound_label}
    \medskip
    \begin{tabular}{c|l|l}
        \hline
         Rank & Sound Belonging & Scene \\ [0.5ex]
         \hline
         \hline
         1 & Doctors &	Hospital \\
         2 & Patients & Ward \\
         3 & Patient’s family & Operation room \\
         4 & Humans	& Hospital corridor \\
         5 & Nurses	& Hospital hall \\
         \hline
    \end{tabular}
    \caption{Top 5 Sound Belongings and Scenes}
    \label{tab:sound_scene}
    \medskip
        \begin{tabular}{c|c|c|c}
        \hline
         Metric & Train & Dev & Combined  \\[0.5ex]
         \hline
         \hline
         $\#$ Utterances & 3337 & 371 & 3707 \\
         $\not\!\circ \;$ $\#$token & 11.14 & 11.19 & 11.14 \\
         $\max$ $\#$token & 48 & 54 & 54 \\
         $\min$ $\#$token & 1 & 1 & 1 \\
         duration(h) & 9.3 & 1 & 10.3 \\
         \hline
    \end{tabular}
    \caption{Data metrics between training and development sets. Token metrics were calculated using the Chinese dataset.}
    \label{tab:data_stats}
\end{table}

The collected dataset was divided into a training and development subset such that the average number of tokens (words) within each subset was roughly equal (see \Cref{tab:data_stats} and \Cref{fig:token_dist}).

\section{Experiments}

The baseline with instructions on how to obtain the proposed dataset and how to run experiments can be found online\footnote{www.github.com/richermans/AudioCaption}.

\paragraph*{Data preprocessing}

Since this paper lays foundation for a new task, different feature types were investigated. Similar features used for tasks such as automatic speech recognition (ASR) and audio event detection (AED) were extracted. We adopted standard log-melspectrogram (AED), as well as filterbank (ASR) features as our baseline (as seen in \Cref{tab:feature_stats}).

\begin{figure}[tb]
	\centering
	\includegraphics[width=0.85\columnwidth]{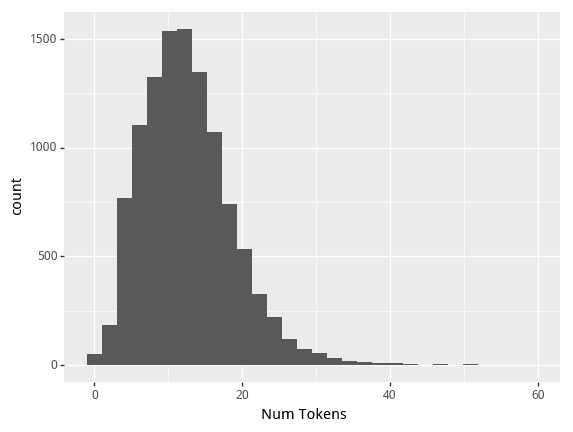}
	\caption{Distribution of sentence length by number of tokens}
	\label{fig:token_dist}
\end{figure}

\begin{table}[htbp]
    \centering
    \small
    \begin{tabular}{c|c|c|c}
        \hline
         Featurename &  Window & Shift & Dimension \\
         \hline\hline
         Lms128 & 40 & 20 & 128 \tabularnewline
         Fbank64 & 25 & 10 & 64 \tabularnewline
         \hline
    \end{tabular}
    \caption{Detailed information on feature extraction parameters. Lms128 stands for logmelspectrogram, fbank64 for filterbank features. Window and Shift values are given in milliseconds (ms).}
    \label{tab:feature_stats}
\end{table}

Regarding the AED features, 128 dimensional logmelspectra \cite{Choi2016,Hershey2016,Xu2017,Kong2018} were extracted. Here, a single frame is extracted every 20ms with a window size of 40ms (\Cref{tab:feature_stats}). Moreover standard 64 dimensional ASR filterbank features were extracted every 10ms with a window size of 20ms. The feature value range is normalized by the mean and standard deviation of the training dataset.
The dataset was labelled in written Mandarin Chinese, a language in which words are not separated by white spaces. Therefore, a tokenizer is needed to split a given sentence into its semantic components (tokens). The Stanfords NLP toolkit \cite{manning-EtAl:2014:P14-5} was utilized to extract tokens (here we opted for words) for each sentence. Commas, dots and other Chinese specific calligraphy symbols were removed. Regarding the translated English sentences, the same preprocessing pipeline was utilized. The resulting token length distribution can be seen in \Cref{tab:data_stats} and \Cref{fig:token_dist}.

\paragraph*{Model}

\begin{figure}[htp]
  \centering
  \includegraphics[width=0.85\columnwidth]{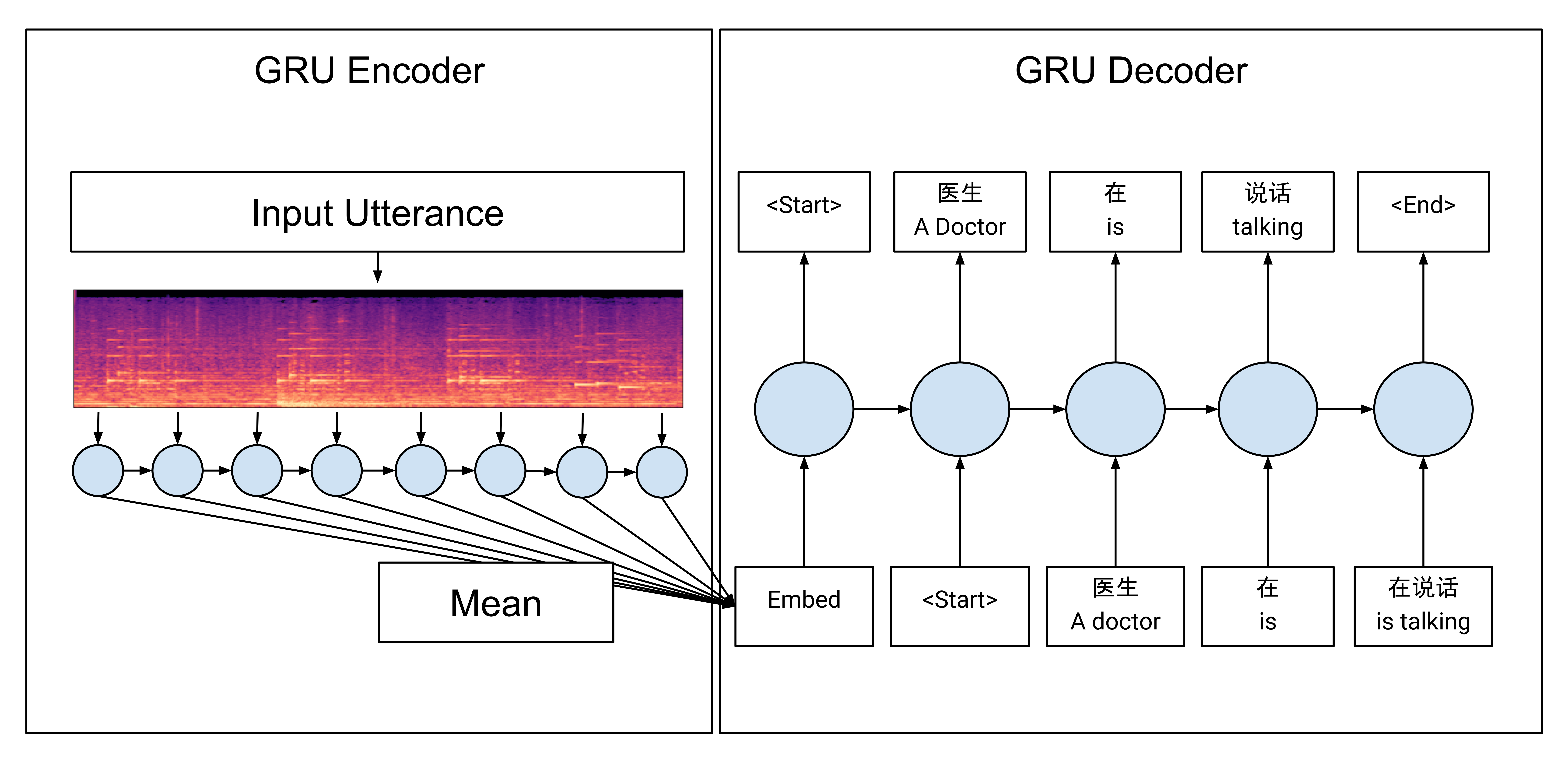}
  \caption{Proposed GRU Encoder-Decoder Model. Circles represent the hidden state of the recurrent model. An embedding $\mathbf{u}$ for a input utterance is extracted by using mean reduction over all time steps.}
  \label{fig:encoder_decoder}
\end{figure}

The objective of this task is to generate reasonable sentences from fed-in audio utterances.The baseline model proposed for this task adopts an encoder-decoder approach, similar to that of end-to-end automatic speech recognition \cite{zhc00-chen-icassp19,zhc00-chen-icassp17-e2e} and image captioning \cite{Vinyals_2015,Karpathy:2017:DVA:3069214.3069250} tasks. The encoder outputs a single fixed dimensional vector $\mathbf{u}$ for each utterance. This fixed dimensional vector is then concatenated with the source sentence and fed to the decoder (see \Cref{fig:encoder_decoder}). The decoder produces an output word for each input word. 

The encoder model consists of a single layer gated recurrent unit (GRU) model. We fixed the embedding size to be a 256 dimensional vector and the hidden size to be 512. The mean of all timestep outputs is used as the utterance representation (denoted as $\mathbf{u}$) (see \Cref{fig:encoder_decoder}).

Furthermore, the decoder model also uses a single layer GRU structure, but applies its dropout in addition to the word embedding output.

\begin{equation}\label{eq:nll}
\mathcal{L}(S | \mathbf{u}; \theta) = - \sum_{t=1}^T \log p_t(S_t | \mathbf{u}; \theta)
\end{equation}

The training loss is the negative log likelihood of the correct word ($S_t$) at each timestep $t$ [see \Cref{eq:nll}] given the model parameters $\theta$ as well as the embedding vector $\mathbf{u}$ .

Training is run for at most 20 epochs, whereas the epoch generating the lowest perplexity on the training set is chosen for sampling. Sampling is done greedily, by inserting the most likely output as the input of the next timestep and done up until 50 words are generated or an end token is seen.

\paragraph*{Evaluation}
Since this dataset uses Chinese labels, the choice of evaluation metrics is limited to language agnostic methods such as BLEU \cite{Papineni:2002:BMA:1073083.1073135}. BLEU can be regarded as a weighted $N$-gram. In this work, we provide scores up to 4-gram. As seen in \Cref{tab:data_stats}, some utterances contain less than four tokens, which leads to an unstable BLEU score. In cases the system generates a hypothesis $<N$, we use method 1 of chancherry smoothing \cite{chen2014systematic}. The results in \Cref{tab:results} are evaluated on the held out development set.

\begin{table}[htpb]
\centering
\small

\begin{tabular}{|r|c|c|c|c|c|}
\hline
\multirow{2}{*}{Feature} & \multirow{2}{*}{Language} & \multicolumn{4}{c|}{$\text{BLEU}_N$}\tabularnewline
\cline{3-6}
 & & 1 & 2 & 3 & 4 \tabularnewline
\hline
\hline
Human &\multirow{3}{*}{English} & 0.376 & 0.196 & 0.121 & 0.086\tabularnewline
Fbank64 && 0.375 & 0.188 & 0.097 & 0.060\tabularnewline
Lms128 && 0.367 & 0.198 & 0.109 & 0.069 \tabularnewline
\hline
Human & \multirow{3}{*}{Chinese} & 0.369 & 0.201 & 0.122 & 0.088\tabularnewline
Fbank64 && 0.339 & 0.161 & 0.088 & 0.058 \tabularnewline
Lms128 && 0.369 & 0.192 & 0.113 & 0.076\tabularnewline

\hline
\end{tabular}
\caption{Evaluation results on the development set. $\text{BLEU}_N$ refers to the BLEU score up until $N$ grams. Each hypothesis sentence is evaluated against two references. The average of all 3 references being treated as hypothesis is displayed as the human score.}
\label{tab:results}
\end{table}

\begin{equation}
\begin{split}
    \text{humanscore}(N) &= \frac{1}{D} \sum_{d=1}^D \frac{1}{R} \sum_{r=1}^R \sum_{q\neq r} \text{BLEU}_{N} (\text{Ref}_{r}, \text{Ref}_{q}) \\
    \text{modelscore}(N) &= \frac{1}{D} \sum_{d=1}^D \frac{1}{R} \sum_{r=1}^R \sum_{q\neq r} \text{BLEU}_{N} (\text{Hyp}, \text{Ref}_{q})
    \label{eq:score}
    \end{split}
\end{equation}

In order to compute the human score, we treat one sentence ($\text{Ref}_{r}$) as hypothesis and the other two ($\text{Ref}_{q}$) as being references. Then we average the scores generated by treating each sentence as hypothesis (as seen in \Cref{eq:score}). This procedure is unfortunately biased towards model generated sentence, since its output has three corresponding references. For adequate comparisons between both outputs, we only compare the model hypothesis with $Q=R-1$ references, where $R=3$ and $D$ is the number of utterances in the dataset. The final BLEU score is then computed as the average of all per utterance BLEU scores.
The performance in \Cref{tab:results} indicates that AED features are more suitable for the audio caption task across both languages. Our baseline model exhibits close to human performance, in both Chinese and English. Therefore, we further investigated the BLEU score reliability by reevaluating the generated sentences by human raters. 

\paragraph*{Human Evaluation} We invited eight native Mandarin speakers to evaluate our Chinese hypothesis by scoring the four captions (one hypothesis from our best model and three human references) from 1 to 4 by its extent of usefulness: 1 stands for the least and 4 indicates the most. Our model averaged 1.89 and humans' captions achieved a score of 2.88, with a two-tailed $t$-test yielding $p$ \textless.001. Fleiss' Kappa score \cite{fleiss1971measuring} was calculated for the eight human evaluators and results indicated that for Hypothesis, almost perfect agreement ($\kappa$ = 0.82) was achieved on `1'. It is quite agreeable that the generated captions are not useful compared to human captions. However, little agreement on the three references was found (average $\kappa$ = 0.28), meaning that even raters agree the human captions are more useful, the extent is subject to preference. This contradictory result to the BLEU score shows that humans are still more reliable and accurate in captioning audio events, thus more attempts are encouraged on this task. 



\section{Discussion}
\label{sec:discussion}
Surprisingly, the baseline results shown in this work achieve scores nearly equal to human scores with BLEU as our criteria, regardless of the language adopted (English or Chinese). We provide a few possible explanations for this behaviour. First, since the dataset is relatively small and domain-specific, many labels can describe multiple utterances. As an example, since the word sound (声) appears frequently, generated sentences are likely to include this word. Second, because of the large variety of words used in Chinese to describe mundane things (e.g., the English word talk can be one of 交谈，说话，聊天，聊，讲，谈论，讨论，商量), the human captions have great variance thus the computed BLEU score is relatively low with humans. 

Additionally, we found that the captions generated by our baseline are consistently more repetitive compared to humans. For human references, there are 1236 unique words (averaged while for our hypothesis, only 193 unique words occurred), indicating that humans can generate sentences with a variety of lexicon, while more repetitive words were generated by our model. Statistics on unique sentences mirrored this observation: 1060 unique sentences from human raters while merely 106 from our model. The overlapping unigrams between hypothesis and reference (80 percent overlap for top 10 unigrams and 83 percent for top 30) showed that the model tends to produce the most possible words. As the current scene is limited to hospital, the model has a bigger chance in modelling relevant captions. However under further evaluation by humans, our model reveals its defect. A few instances can be seen in the box below.\\ 

\noindent\fbox{
\footnotesize{
\parbox{\columnwidth}{
\textbf{\textit{Hyp (Score 4 Most Useful):医生在病房里医生和病人的对话声\\
The doctor's conversation with the patient in the ward.\\}}
Ref 1:病人和医生对话 (Score 2) \\
Dialogue between patients and doctors\\
Ref 2:在病房里，有说话声 (Score 1) \\
In the ward, there was a voice.\\
Ref 3:病人和护士的对话 (Score 3) \\
Dialogue between patients and nurses\\
\vspace{3mm}
\textbf{\textit{Hyp (Score 3 Medium Useful):医生和病人家属在说话有人的对话声\\
The doctor and the patient's family are talking.\\}}
Ref 1:一位医生与一位女士交谈 (Score 2) \\
A doctor talked to a lady.\\
Ref 2:病患询问医生派出所的位置 (Score 1) \\
The patient inquired about the location of the doctor's police station.\\ 
Ref 3:大厅广播声，脚步声，医生和病人对话声，大厅嘈杂声 (Score 4) \\
The hall was broadcasted, footsteps, doctors and patients chatting, the hall was noisy.\\
\vspace{3mm}
\textbf{\textit{Hyp (Score 1 Not Useful) 医生和病人的对话声\\
Dialogue between doctors and patients\\}}
Ref 1:一个医生和另一个正在打电话的医生交谈，说话声，走路声，吵杂声 (Score 4) \\
A doctor talked to another doctor on the phone, speaking, walking, and noisy.\\
Ref 2:两个医生在医院对话，但并不是说同一件事情 (Score 3) \\
The two doctors talked in the hospital, but not the same thing.\\
Ref 3:男医生说话，另一名男医生电话交流 (Score 2) \\
Male doctor speaks, another male doctor exchanges by telephone.\\
}}
}

\section{Conclusion}
\label{sec:conclusion}

In this paper we endeavored to enhance machine perception of audio events, and to minimize the gap between audio and image research. The aim is to automatically generate captions for audio events. We collected 3710 video clips on hospital scene, with each clip containing more than one sound classes. We established a new way of subjective labelling to enable automatic audio caption. Each 10s clip is entitled to three human-labelled captions. An encoder-decoder model was conducted as our baseline and experiment results show that the best results were achieved with logmelspectrogram input features. To our surprise, this model`s BLEU score is slightly lower than the human one. However, this might not be the best way to evaluate Chinese captions: human evaluation showed that our model is still not very accurate. We could generate grammatically-correct and data-related captions but the model tends to produce repetitive words and sentences. This behaviour is likely a result of the limited data diversity. In future work we would like to extend the dataset to all five provided scenes.

\vfill\pagebreak

\bibliographystyle{IEEEbib}
\bibliography{refs}

\end{CJK*}
\end{document}